\documentclass[pdflatex,sn-nature]{sn-jnl}

\usepackage{graphicx}%
\usepackage{multirow}%
\usepackage{amsmath,amssymb,amsfonts}%
\usepackage{amsthm}%
\usepackage{mathrsfs}%
\usepackage[title]{appendix}%
\usepackage{xcolor}%
\usepackage{textcomp}%
\usepackage{manyfoot}%
\usepackage{booktabs}%
\usepackage{algorithm}%
\usepackage{algorithmicx}%
\usepackage{algpseudocode}%
\usepackage{listings}%
\usepackage{subcaption}%
\usepackage[mathlines]{lineno}%

\unnumbered
\newcommand{\ket}[1]{\ensuremath{\left| {#1} \right>}}
\newcommand{\bra}[1]{\ensuremath{\left< {#1} \right|}}

\begin{document}
%
%
\title[Article Title]{Observation of a Topological Berry Phase with a Single Phonon in an Ion Microtrap Array}


\author*[1,2,a]{\fnm{Justin F.} \sur{Niedermeyer}}\email{justin.niedermeyer@colorado.edu}

\author[1,2,b]{\fnm{Nathan K.} \sur{Lysne}}

\author[1,2]{\fnm{Katherine C.} \sur{McCormick}}

\author[1,2,c]{\fnm{Jonas} \sur{Keller}}

\author[3]{\fnm{Craig W.} \sur{Hogle}}

\author[3]{\fnm{Matthew G.} \sur{Blain}}

\author[3]{\fnm{Edwin J.} \sur{Heller}}

\author[4,d]{\fnm{Roman} \sur{Schmied}}

\author[1,e]{\fnm{Robert} \sur{J\"ordens}}

\author[1,2,a]{\fnm{Susanna L.} \sur{Todaro}}

\author[1,d]{\fnm{David J.} \sur{Wineland}}

\author[1]{\fnm{Andrew C.} \sur{Wilson}}

\author[1]{\fnm{Daniel H.} \sur{Slichter}}

\author*[1]{\fnm{Dietrich} \sur{Leibfried}}\email{dietrich.leibfried@nist.gov}

\affil[1]{\orgname{National Institute of Standards and Technology}, \orgaddress{\city{Boulder}, \state{CO} \postcode{80305}, \country{United States}}}

\affil[2]{\orgdiv{Department of Physics}, \orgname{University of Colorado Boulder}, \orgaddress{\city{Boulder}, \state{CO} \postcode{80309}, \country{United States}}}

\affil[3]{\orgname{Sandia National Laboratories}, \orgaddress{\city{Albuquerque}, \state{NM} \postcode{87123}, \country{United States}}}

\affil[4]{\orgdiv{Department of Physics}, \orgname{University of Basel}, \postcode{CH-4056} \city{Basel}, \country{Switzerland}}

\affil[a]{\orgdiv{Present Address:} \orgname{Oxford Ionics}, \city{Oxford}, \postcode{OX5 1GN}, \country{United Kingdom}}

\affil[b]{\orgdiv{Present Address:} \orgname{Quantinuum K.K.}, \city{Chiyoda-ku}, \state{Tokyo} \postcode{100-0004}, \country{Japan}}

\affil[c]{\orgdiv{Present Address:} \orgname{Physikalisch-Technische Bundesanstalt}, \postcode{D-38116} \city{Braunschweig}, \country{Germany}}

\affil[d]{\orgdiv{Present Address:} \orgname{Viewpointsystem GmbH}, \postcode{A-1010} \city{Vienna}, \country{Austria}}

\affil[c]{\orgdiv{Present Address:} \orgname{QUARTIQ GmbH}, \postcode{D-12489} \city{Berlin}, \country{Germany}}

\affil[d]{\orgdiv{Present Address:} \orgname{Department of Physics, University of Oregon}, \postcode{97403-1274} \city{Eugene, OR}, \country{United States}}

\maketitle
%
\section{Summary Paragraph}\label{sec:alt-summ-para}

Controlled quantum mechanical motion of trapped atomic ions can be used to simulate and explore collective quantum phenomena~\cite{porras_2004, porras_2004b} and to process quantum information~\cite{cirac_1995}. Groups of cold atomic ions in an externally applied trapping potential self-organize into ``Coulomb crystals'' due to their mutual electrostatic repulsion~\cite{diedrich_1987,wineland_1987,gilbert_1988}. The motion of the ions in these crystals is strongly coupled, and the eigenmodes of motion all involve multiple ions. While this enables studies of many-body physics~\cite{britton_engineered_2012, bohnet_quantum_2016, zhang_2017}, it limits the flexibility and tunability of the system as a quantum platform. Here, we demonstrate an array of trapped ions in individual trapping sites whose motional modes can be controllably coupled and decoupled by tuning the local applied confining potential for each ion~\cite{cirac_scalable_2000,schmied_optimal_2009, shi_topological_2013}. We show that a single motional quantum, or phonon, can be coherently shared among two or three ions confined at the vertices of an equilateral triangle 30 \textmu m on a side~\cite{mielenz_arrays_2016, hakelberg_interference_2019}. We can adiabatically tune the ion participation in the motional modes around a closed contour in configuration space, observing that the single-phonon wavefunction acquires a topological Berry phase~\cite{berry_phase_1984} if the contour encircles a conical intersection of motional eigenvalue surfaces. We observe this phase by single-phonon interference and study its breakdown as the motional mode tuning becomes non-adiabiatic. Our results show that precise, individual quantum control of ion motion in a two-dimensional array can provide unique access to quantum multi-body effects.
\section{Introduction}\label{sec:introduction}
\begin{figure*}[htb!]
\centering
    \begin{subfigure}{\textwidth}
        \centering
        \includegraphics[width=1.\textwidth]{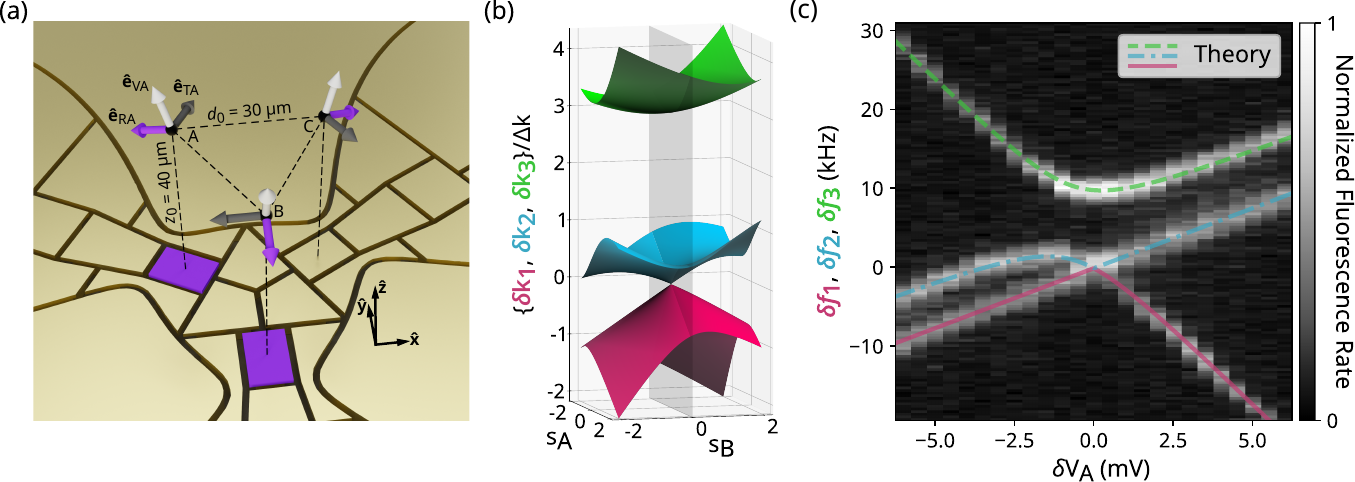}
        \phantomsubcaption
        \label{fig:trap}
    \end{subfigure}
    \begin{subfigure}{0.\textwidth}
        \phantomsubcaption
        \label{fig:mode-surfaces}
    \end{subfigure}
    \begin{subfigure}{0.\textwidth}
        \phantomsubcaption
        \label{fig:spectrum}
    \end{subfigure}
    \caption{(a) Rendering of the triangular 2D ion microtrap array. Electrodes are shown in gold, gaps in black. Black spheres labeled A, B, and C mark the positions of the potential wells where $^9$Be$^+$ ions are trapped. Colored arrows labeled $\hat{\mathbf{e}}_{\rm RA}$ (purple), $\hat{\mathbf{e}}_{\rm VA}$, (white), $\hat{\mathbf{e}}_{\rm TA}$ (gray) indicate the radial, vertical, and tangential principal axis directions in site A, respectively. The principal axes in the other sites are shown but not labeled. The electrodes closest to site A and site B (purple) are used to fine-tune the site curvatures. (b) The radial motion eigenvalue surfaces vs. tuning parameters $s_A$ and $s_B$. The lower two manifolds (blue and red) form a conical intersection at $s_A = s_B = 0$. The semitransparent gray plane indicates a cut through the surfaces along $s_B=0$ where the eigenmode frequencies (related to the curvature eigenvalues by $\delta f_j = \tfrac{1}{2 \pi}\delta k_j/(2 m\, \omega_R)$) were experimentally measured, with data shown in (c). The observed normalized fluorescence rate is shown as a function of the curvature tuning voltage $\delta V_A$ and the difference $\delta f$ of the excitation frequency from the eigenfrequency conical intersection $\omega_R/(2 \pi)$. Larger motional excitation leads to higher normalized fluorescence rates. The colored lines are the best fit to theory with eigenfrequencies colored in analogy to Fig. \ref{fig:mode-surfaces} (for details see \nameref{sec:methods}). }
\end{figure*}

The quantized shared motion of trapped ions is essential to their use for quantum simulation~\cite{porras_2004,porras_2004b}, quantum information processing~\cite{cirac_1995}, and quantum logic spectroscopy~\cite{brewer_2019, chou_2017_molecule, micke_2020_HCI}. Almost all work in these fields uses traps that create approximately harmonic potentials with a single minimum. Depending on the number of ions and the geometry of the trapping potential, the ions self-organize around this minimum in one-dimensional strings~\cite{raizen_1992} or higher-dimensional arrangements~\cite{diedrich_1987,wineland_1987,gilbert_1988,wang_2020,ivory_2020,xie_openendcap_2021,kiesenhofer_2023, guo_siteresolved_2024}. External fields are then applied to control the interactions of the ions bound in that potential by driving their coupled modes of motion; such systems have been used to perform a variety of quantum simulations~\cite{britton_engineered_2012,bohnet_quantum_2016,zhang_2017,li_2023,hainzer_2024}. In two dimensions (2D) and for sufficiently large numbers of ions, triangular lattices are typically formed in the crystal centers. Lattice defects arising from a variety of technical sources can produce many different metastable crystal configurations that are not necessarily at the lowest energy but are stable over long timescales~\cite{kiesenhofer_2023}. This hard-to-control variation in larger self-organized crystals leads to corresponding variation in the ion-ion interactions. Precise control over couplings between ion pairs is further complicated by unequal participation of the ions in each of the motional modes, posing a substantial challenge for tuning interaction strengths as the number of ions increases.\\
\\
In principle, arbitrary 2D geometries of ions, including regular lattices, quasi-periodic tilings, non-Euclidean lattices, and tunable defects or dislocations, can be realized by trapping ions individually in a set of separate, independently controllable potential wells in a microtrap array~\cite{chiaverini_2008,schmied_optimal_2009}. Such arrangements can create inherent symmetries and topological features of dynamically tunable ion interactions not accessible to self-organised ion crystals that occupy a single potential well. To observe coherent interactions, the wells need to be close enough to each other that the Coulomb coupling rate is substantially larger than all motional decoherence rates in the system. Previous demonstrations of microtrap arrays have shown the ability to load and deterministically rearrange ions among wells~\cite{holz_2d_2020, palani_highfidelity_2023}, but the creation of individually controllable wells requires electrodes in close proximity to the ions, such that the motional heating rates were equal to or larger than the motional coupling rate between wells~\cite{mielenz_arrays_2016,hakelberg_interference_2019}.\\
\\
Here, we demonstrate quantum-coherent motional coupling and the direct observation of a topological Berry phase in a fully tunable 2D microtrap array consisting of three ions arranged in an equilateral triangle (Fig. \ref{fig:trap}). We prepare all nine modes of motion close to their ground states and inject a single phonon to create an entangled Bell state of the coupled eigenmodes.  
By dynamically tuning the couplings between motional modes, we can continuously deform the character of the single-phonon motional state to encircle a conical intersection in the motional eigenvalue landscape, evidenced by a topological Berry phase~\cite{berry_phase_1984} visible in single-phonon interference experiments. To the best of our knowledge, Berry phases from conical intersections have previously only been inferred from the observation of node lines in wavepacket distributions in quantum simulations~\cite{wang_2023,valahu_2023,whitlow_2023}. \\ 
\\
We now turn to the theory of coupled motional modes for an array of three potential wells, considering here the specific case of an equilateral triangle. Given a set of nearly equal potential curvatures (second spatial derivatives of the potential) along equivalent principal axes $\hat{\mathbf{e}}_{RA}$, $\hat{\mathbf{e}}_{RB} $ and $\hat{\mathbf{e}}_{RC} $ of each of the three sites labeled A, B, C ($\hat{\mathbf{e}}_{RA}$ is shown in Fig. \ref{fig:trap}), all physical details of the ions, the trap, and the Coulomb interactions can be represented by three unitless 3-by-3 matrices and five quantities that fully specify the curvature matrix (Hessian) $\mathbf{H}$ and determine the eigenvalues and eigenvectors of the coupled ion motion (see \nameref{sec:methods} for more details). In the basis of the uncoupled oscillators with products of number states $\ket{k l m}_{ABC} =\ket{k}_A \ket{l}_B \ket{m}_C$ the Hessian is given by
\begin{equation}\label{Eq:HesMat}
 \mathbf{H}=
k_{\rm offs}
\left(
\begin{array}{ccc}
  1&0&0 \\
  0&1&0 \\
  0&0&1   
\end{array}
\right)
+
\Delta k
\left(
\begin{array}{ccc}
  1&1&1 \\
  1&1&1 \\
  1&1&1   
\end{array}
\right)
+
\Delta k
\left(
\begin{array}{ccc}
    s_A+\alpha s_B&0&0 \\
  0&s_B+\alpha s_A&0 \\
  0&0&\alpha(s_A+s_B) 
\end{array}
\right). 
\end{equation}
The curvature $k_{\rm offs}$ expresses equal curvature of the site potentials along a set of equivalent principal axes ($\hat{\mathbf{e}}_{RA}$, $\hat{\mathbf{e}}_{RB} $ and $\hat{\mathbf{e}}_{RC} $, see Fig. \ref{fig:trap}). It multiplies the identity matrix, and consequently any three-component vector is an eigenvector of this part with eigenvalue $k_{\rm offs}$ that offsets the eigenvalues of the remainder of  $\mathbf{H}$. The curvature $\Delta k$ characterizes the Coulomb interaction and multiplies the interaction matrix with all elements equal to one, which reflects the triangular symmetry of the system that renders the Coulomb interaction between any pair of ions identical. In our implementation $k_{\rm offs} \gg \Delta k >0$ and the interaction matrix has one eigenvalue $\delta k_3=3 \Delta k$ with eigenvector $\mathbf{c}^{(3)}=(c_A^{(3)},c_B^{(3)},c_C^{(3)})^T=1/\sqrt{3}(1,1,1)^T$ and eigenvalues $\delta k_{1,2}=0$ for any two vectors that are mutually orthogonal and orthogonal to $\mathbf{c}^{(3)}$. One choice is $\mathbf{c}^{(1)}=1/\sqrt{2}(0,1,-1)^T, \mathbf{c}^{(2)}=1/\sqrt{6}(2,-1,-1)^T$. 
The unitless quantities $s_A, s_B$ and $\alpha$ describe tuning factors added to the site curvatures that we can realize by applying suitable potentials to the electrodes closest to the sites labeled A and B, shown in purple in Fig. \ref{fig:trap}. The ion in site A does not participate in eigenmode $\mathbf{c}^{(1)}$. While $s_B=0$, $\mathbf{c}^{(1)}$ remains an eigenvector of $\mathbf{H}$ for $s_A \neq 0$ with an eigenvalue $k_{\rm offs}+\Delta k~ \alpha~ s_A$ that is linear in $s_A$.\\
\\
The normalized differences $\delta k_j/\Delta k$ of the eigenvalues of $\mathbf{H}$ from $k_{\rm offs}$ as a function of general $s_A$ and $s_B$ are shown in Fig. \ref{fig:mode-surfaces}. Two of the eigenvalue surfaces (shown in pink and blue) meet in a conical intersection at $s_A= s_B =0$ with coalescing eigenvalues $k_{\rm offs}$ while the third surface (green) maintains a substantial gap $\geq 3 \Delta k$ to the other two surfaces. The conical intersection is a consequence of the symmetry of the array. We can prepare an eigenstate or a superposition of eigenstates of the motion and then tune the local curvatures, and thereby the Hamiltonian of the coupled ion motion, by changing $s_A$ and $s_B$. If this tuning is sufficiently slow, the adiabatic theorem applies and initial eigenstates of the coupled motion remain eigenstates throughout. Eigenstates such as those with eigenvalues on the lower two surfaces acquire a topological (or Berry) phase of $\pi$ when $s_A$ and $s_B$ are adiabatically tuned along a closed path that encloses the conical intersection at $s_A = s_B = 0$~\cite{berry_phase_1984}. If the conical intersection is not enclosed, for example if the initial state is an eigenstate with eigenvalue on the top surface (green), or if a path on the lower two surfaces does not enclose the conical intersection, the Berry phase is zero. The Berry phase acquired in either case is independent of the shape of the path and the duration required to move through it as long as the process remains sufficiently adiabatic. 
\section{The Triangular Ion Microtrap Array}\label{sec:trap}
A view of the 2D ion microtrap array is shown in Fig. \ref{fig:trap}. The confining potential is produced by a single rf electrode with a hexagonal outer shape and a ``cloverleaf'' shaped cutout with threefold symmetry that is segmented into 30 control electrodes~\cite{lysne_individual_2024,niedermeyer_thesis_2025}. The trap is microfabricated in a multilayer process with a top electrode layer made of aluminum coated with evaporated gold and buried electrical routing layers. Gaps in the electrodes are approximately that of the metal thickness {($\sim1.4\, \mu$m)} and the insulating dielectric (SiO$_2$) below the electrodes is hidden from line-of-sight to ion positions.  The metal layer immediately below the top metal/electrode layer is grounded in the electrode gap regions~\cite{maunz_high_2016}. We apply a potential to the rf electrode with an amplitude of $\sim$42 V oscillating at 121.1 MHz to create three potential wells on the corners of an equilateral triangle with side length $d_0=30$ \textmu m that are $z_0=40$ \textmu m from the electrode surface. We trap a single $^9$Be$^+$ ion in each potential well.\\
\\
By applying linear combinations of voltages to the 30 control electrodes, we can tune the electric field vector at the position of a given ion without affecting the curvature tensor (and vice versa), and without appreciably changing the curvatures or fields at the positions of the other two ions. Site-specific applied electric fields can be used to compensate stray electric fields and minimize ion micro-motion~\cite{berkeland_1998} or for individual addressing~\cite{lysne_individual_2024}. Site-specific curvature tuning is used to compensate for stray curvatures and to adjust the principal axis orientations and confinement strengths in each well. The nine principal axes, three per well, adhere to the $120^\circ$ rotational symmetry of the rf electrode, as shown in Fig.~\ref{fig:trap}. In one of these wells, a single $^9$Be$^+$ ion has motional frequencies $\left\{\omega_{R}, \omega_{V}, \omega_{T}\right\}\simeq2\pi\times\left\{3.9 , 6.9, 10.8\right\}$ MHz, respectively along the principal axes, which we name radial (purple), vertical (white), and tangential (grey). The curvatures of all wells are fine tuned along their radial directions by changing the potentials on the electrodes closest to the ion in the well (see Fig.~\ref{fig:trap}). The motional frequency of radial eigenmode $j$ is connected to its eigenmode curvature by
\begin{eqnarray}\label{Eq:DelOme}
\omega_j = \sqrt{\frac{k_{\rm offs}+\delta k_j}{m}} &\approx& \omega_R+\delta \omega_j,\nonumber \\
\omega_{R}=\sqrt{\frac{k_{\rm offs}}{m}}, && \delta \omega_j = \frac{\delta k_j}{2 m\,\omega_R},
\end{eqnarray}
where $m$ is the mass of a $^9$Be$^+$ ion. The rate of change of the scaled curvatures with the applied potentials is $\delta s_A/\delta V_A=\delta s_B/\delta V_B \approx$ 1.2 mV$^{-1}$. By symmetry the curvatures in sites B and C change at equal amounts $\alpha~ \delta s_A/\delta V_A$ with $\alpha \approx -0.383$ and likewise for A and C when $\delta V_B$ is changed (see \nameref{sec:methods}).\\  
\\
We apply an approximately 0.5 mT magnetic field to lift the degeneracies in the ${^2S_{1/2}}$ ground state manifold of $^9\text{Be}^+$, where the $\left|\downarrow\right\rangle\equiv\left|F=2,m_F=-2\right\rangle$ and the $\left|\uparrow\right\rangle\equiv\left|F=1,m_F=-1\right\rangle$ hyperfine states are split by approximately 1.26 GHz. We can drive transitions between all hyperfine ground states by applying a microwave magnetic field resonant with their frequency differences using a half-wave antenna embedded in the ground plane of the trap chip $\approx$ 380 \textmu m away from the trap center~\cite{lysne_individual_2024,niedermeyer_thesis_2025}. An elliptically shaped laser beam with a wavelength of $\approx$ 313 nm illuminates all three ions and is used for Doppler cooling and qubit readout via state-dependent fluorescence on the cycling $\left|\downarrow\right\rangle\leftrightarrow\,{^2P_{3/2}}\ket{F=3, m_F =-3}$ transition~\cite{wineland_experimental_1998}. We perform ground state cooling of all motional modes, as well as motional state preparation and analysis of the three ions, using stimulated Raman transitions driven by a pair of counter-propagating, tightly focused 313 nm laser beams detuned $\approx 80$ GHz from the ${^2S_{1/2}}\leftrightarrow{^2P_{1/2}}$ transition~\cite{monroe_1995_raman_cooling, wineland_experimental_1998}. We achieve an average occupation $\bar{n} < 0.05$  for all nine modes after ground state cooling. Due to the geometry of the vacuum chamber, these Raman beams can either illuminate the ions in sites A and B (``ion A'' and ``ion B'') simultaneously or just a single ion in site C (``ion C''). \\
\\
A helium-gas-vibration-isolated closed-cycle cryostat~\cite{todaro_scalable_2020} cools the trap and its copper enclosure to $\sim$3.5 K, providing cryopumping of residual background gas and reduced ion motional heating rates. Ion lifetimes in the trap are typically many weeks (chiefly limited by user error), and the motional heating rates at this temperature are measured to be approximately $\left
\{\dot{\overline{n}}_{Rj}, \dot{\overline{n}}_{Vj}, \dot{\overline{n}}_{Tj}\right\} \simeq \left\{30, 14, 2\right\}$ phonons/s for the three modes of a single ion confined in any of the three sites ($j\in\{ A,B,C\}$), much lower than the inter-ion coupling rates. 
\section{Static Interaction Tuning}\label{sec:static-tuning}
When $|s_A|$ and $|s_B|$ are close to zero, the Coulomb interaction couples the radial motion of the ions into three collective normal modes $\mathbf{c}_1, \mathbf{c}_2, \mathbf{c}_3$ with scaled curvature eigenvalues $\delta k_1, \delta k_2, \delta k_3$ as shown in Fig.~\ref{fig:mode-surfaces}. We experimentally characterize a cut through the eigenfrequency surfaces along $s_B=0$ (corresponding to the semitransparent plane in Fig.~\ref{fig:mode-surfaces}) by tuning the potential $\delta V_A$ applied to the electrode nearest to site A (see Fig.~\ref{fig:trap}). The Raman beams are aligned with ions A and B and all modes of motion are cooled to near their ground states. Then, the hyperfine states are optically pumped to ideally prepare $\ket{ \downarrow \downarrow \downarrow}_{ABC}\ket{000}_{123}$, where $\ket{lmn}_{123} = \ket{l}_{1} \ket{m}_{2} \ket{n}_{3}$ is a product of number states in the basis of the three coupled radial eigenmodes (note that the coupled $123$-basis is distinct from the uncoupled $ABC$-basis).\\
\\
We then apply a weak oscillating electric field pulse at the ion positions using the chip-integrated antenna and scan its frequency to probe the motional eigenmode frequencies. When the tone is resonant with one of the normal modes, the mode will be excited into a small coherent state \cite{mccormick_2019_coherent} with a maximal average motional occupation of $\bar{n}_j\approx 1$. We detect the presence or absence of this motional excitation using a Raman red sideband pulse on both ions followed by fluorescence readout of the ion spin state. The normalized  fluorescence of the two ions (which is correlated with the probability that the motion was resonantly excited by the probe electric field) is shown in Fig. \ref{fig:spectrum}. The bright lines trace the normal mode frequencies  in good agreement with the theoretical predictions based on the eigenvalues of Eq. (\ref{Eq:HesMat}) that are shown by colored lines (see \nameref{sec:methods} for details). We observe the degeneracy of two of the modes at $\delta V_A=0$~mV expected from the trap symmetry and the predicted gap to the highest mode frequency. The mode eigenvector $\mathbf{c}_1 $ remains equal to $ 1/\sqrt{2}(0,-1,1)$ throughout with $\delta \omega_1$ increasing linearly with $\delta V_A$, as expected.  Small deviations from the theory are caused by slow drifts in the applied and stray potentials on the order of tens of microvolts 
\begin{figure*}[htb!]
\centering
    \begin{subfigure}{\textwidth}
        \centering
        \makebox[\textwidth][c]{\includegraphics[width=\textwidth]{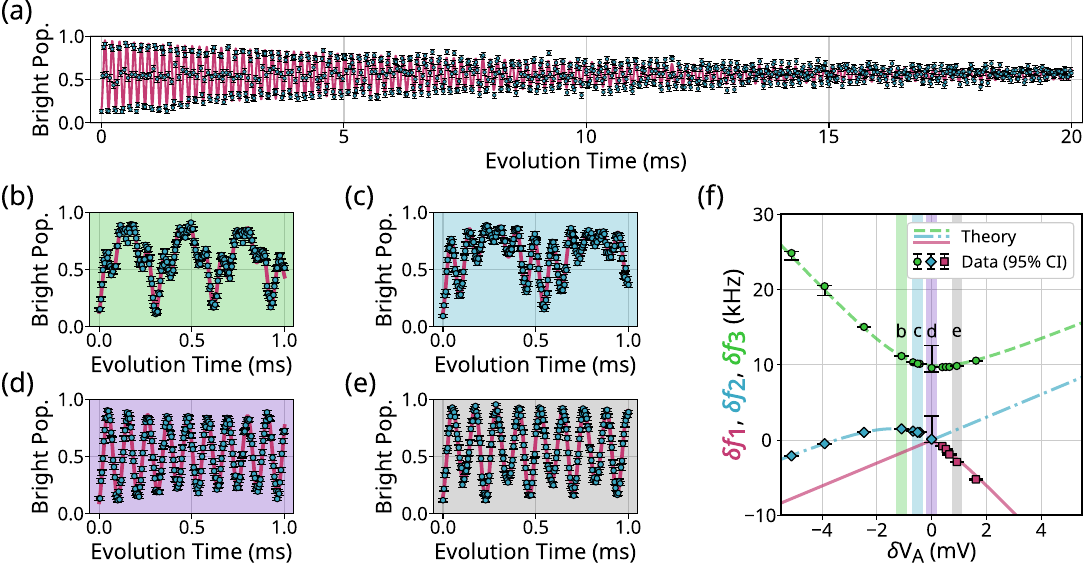}}
        \phantomsubcaption
        \label{fig:exchange-a}
    \end{subfigure}
    \begin{subfigure}{0.\textwidth}
        \phantomsubcaption
        \label{fig:exchange-b}
    \end{subfigure}
    \begin{subfigure}{0.\textwidth}
        \phantomsubcaption
        \label{fig:exchange-c}
    \end{subfigure}
    \begin{subfigure}{0.\textwidth}
        \phantomsubcaption
        \label{fig:exchange-d}
    \end{subfigure}
    \begin{subfigure}{0.\textwidth}
        \phantomsubcaption
        \label{fig:exchange-e}
    \end{subfigure}
    \begin{subfigure}{0.\textwidth}
        \phantomsubcaption
        \label{fig:exchange-f}
    \end{subfigure}
\caption{Single phonon interferences. (a) Probability of finding ion C in $\ket{\downarrow}$ as a function of the evolution time between $\pi$-pulses that inject and remove a single phonon. With ions in sites A and C and tuned on resonance, a single phonon injected into C is coherently exchanged hundreds of times. (b)--(e)  All three sites occupied with the ions in B and C on resonance ($\delta V_B \approx 0$), (b) $\delta V_A =$ -1.1, (c) -0.51, (d) 0.00, and (e) 0.92 mV, respectively.  Solid lines are fits to the data. Error bars represent $1\sigma$ of $P_{\downarrow}$. (f) Frequency differences extracted from single phonon interference in (b)--(e), labeled by matching colors; and from additional experiments with different $\delta V_A$. The frequency differences are plotted relative to the eigenfrequency that changes linearly with $\delta V_A$ and are in good agreement with the theoretical eigenmode frequencies for the parameters found from fitting the data in Fig. \ref{fig:spectrum} (solid, dashed and dashed-dotted lines, for details see \nameref{sec:methods}). Error bars represent 95\% confidence intervals obtained from 5000 bootstrapping trials.}
\end{figure*}
that occur on the time scale of hours.\\
\\
To demonstrate control over the dynamics of a single phonon, ions are trapped in A and C and the Raman beams are moved to only illuminate ion C. With the motional frequencies of the two ions on resonance, the ions are cooled near their ground state of motion and transferred to the $\ket{\uparrow}$ state with a microwave $\pi$ pulse. We then drive the approximate transition $\ket{\uparrow\uparrow}_{AC}\ket{00}_{AC} \rightarrow \ket{\uparrow\downarrow}_{AC}\ket{01}_{AC}$ with a Raman sideband $\pi$ pulse on ion C with a duration of 4 \textmu s. The pulse duration is much shorter than the inverse of the coupling strength between ions, $\Delta \omega_j = \Delta k_j/(2 m\, \omega_R)=2\pi \times (3.299\pm 0.002)$ kHz, and therefore to a good approximation the phonon is added to ion C before it can transfer appreciably to ion A. The spectral width of the Raman sideband pulse spans the frequencies of both radial eigenmodes, and they therefore can be excited simultaneously by the pulse, creating an entangled state of a single phonon distributed over two eigenmodes~\cite{brown_coupled_2011,harlander_2011,wilson_tunable_2014}. By applying another sideband $\pi$ pulse on ion C at $t\geq0$, the probability of finding the phonon at ion C is mapped onto the probability $P_\downarrow$ of the ion spin state being $\ket{\downarrow}$, which is detected with state-dependent fluorescence.  This probability is shown as a function of the delay $t$ between the sideband pulses in Fig. \ref{fig:exchange-a}. The single phonon is coherently exchanged hundreds of times.\\
\\
With ions in all three sites cooled to their ground states of motion and transferred to the $\ket{\uparrow}$ state we can apply a desired $\delta V_A$ and inject a phonon into site C as described above. The evolution of the approximate initial motional state $\ket{\psi\left(0\right)}=\ket{001}_{ABC}$ can be concisely expressed in the coupled mode basis where the three basis states are energy eigenstates and therefore only acquire phases:
\begin{equation}\label{Eq:TimEvo}
    \ket{\psi(t)} = e^{i \omega_R t}\left( c^{(1)}_Ce^{i \delta \omega_1 t} \ket{100}_{123}+c^{(2)}_C e^{i \delta \omega_2 t}\ket{010}_{123}+c^{(3)}_C e^{i \delta \omega_3 t}\ket{001}_{123}\right),
\end{equation}
where the amplitude of the initial state $|001\rangle_{ABC}$ in the $k$-th coupled mode is given by the third eigenvector component $c_C^{(k)}$. For $t>0$ this evolution periodically delocalizes the phonon over several ions and relocalizes it to ion C (see \nameref{sec:methods} for more details). The second $\pi$ pulse on the sideband of ion C at $t\geq0$ maps the probability $P_{001}(t)=|_{ABC}\langle 001 \ket{\Psi(t)}|^2$ onto the probability of changing the state of ion C back to $\ket{\uparrow}$. The probability $P_{\downarrow}=1-P_{\uparrow}=1-P_{001}(t)$ is then read out by state dependent fluorescence. As a function of delay $t$ between the $\pi$ pulses, $P_{001}(t)$ consists of a sum of up to three sinusoids with amplitudes that depend characteristically on the $c_C^{(k)}$ and oscillate at the differences of the eigenmode frequencies $\delta \omega_k-\delta \omega_l,~ k\neq l$ (see \nameref{sec:methods}). A few examples for different $\delta V_A$ are shown in
Figs.~\ref{fig:exchange-b}--\ref{fig:exchange-e}, where a detection of ion C in the bright state $\ket{\downarrow}$ signals that the second $\pi$ pulse has not flipped the spin state and therefore the phonon was not detected in C. A fit to the sum of up to three sinusoids yields the products $|c_C^{(j)}|^2|c_C^{(k)}|^2$ and the frequency differences $\Delta_j=\delta \omega_j-\delta \omega_1$ for $~j\neq 1$. The fitted  $\Delta_j$ for additional $\delta V_A$ are overlaid with symbols over the theoretical eigenmode spectrum in Fig.~\ref{fig:exchange-f}, showing excellent agreement. This demonstrates that the dynamics of a single phonon can be precisely controlled and behave as expected in the microtrap array. We attribute the imperfect contrast in the exchange data primarily to off-resonant excitation of other transitions due to the relatively large spectral width of the deliberately short sideband $\pi$ pulses.
\section{Observation of a topological Berry phase}\label{sec:dynamic-tuning}
At $s_A = s_B = 0$, the lower two eigenmode surfaces (shown in Fig.~\ref{fig:mode-surfaces} in pink and blue) meet at a single point with $\delta k_1 = \delta k_2 = 0$ and form a conical intersection, while points on the highest eigenmode surface (shown in green) have finite values $\delta k_3\geq 3 \Delta k$ for all $s_A$ and $s_B$.
The conical intersection is a topological feature that gives rise to a Berry phase of $\pi$ for motional states that are prepared on the lower two eigenmode surfaces away from $s_A = s_B = 0$ and then ``transported'' around the intersection by adiabatic changes $s_A(t)$ and $s_B(t)$ on a closed path $s_{A,B}(t=0)=s_{A,B}(t=T)$ over the time interval $[0,T]$~\cite{berry_phase_1984}. When a state acquires a Berry phase, this does not change the expectation values of any observable, $\bra{\psi}\hat{O}\ket{\psi}=\bra{\psi}e^{-i \phi_{\rm Berry}}\hat{O}e^{i \phi_{\rm Berry}}\ket{\psi}$, but the phase change can alter the interference between parts of a superposition of states.\\
\\
\begin{figure*}[htb!]
\centering
    \begin{subfigure}{\textwidth}
        \centering
        \includegraphics[width=\textwidth]{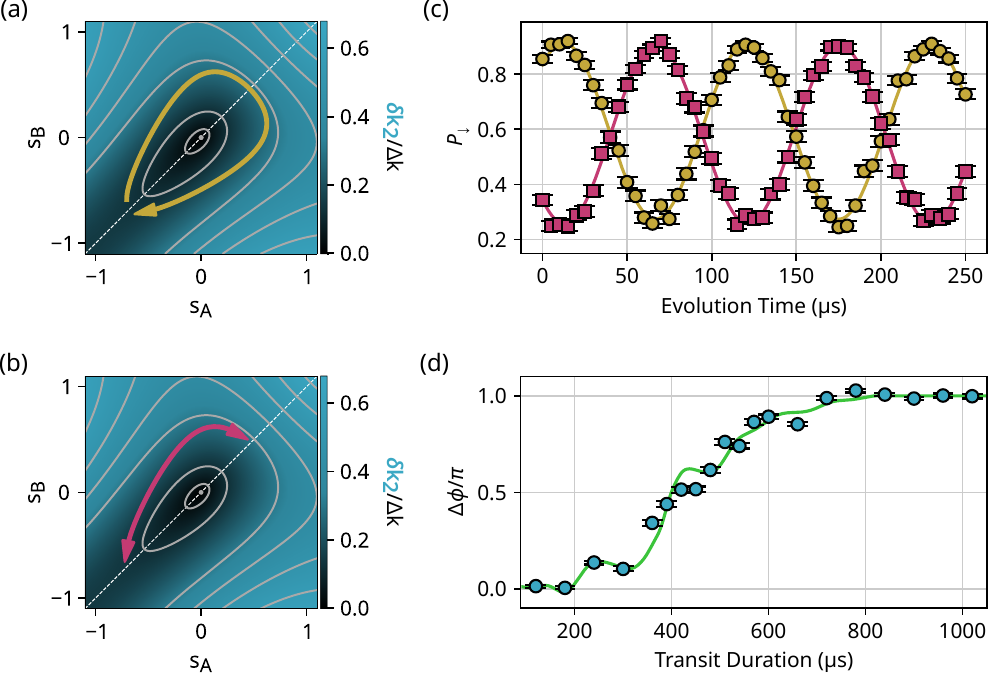}
        \phantomsubcaption
        \label{fig:berry-enclosed-path}
    \end{subfigure}
    \begin{subfigure}{0.\textwidth}
        \phantomsubcaption
        \label{fig:berry-non-enclosed-path}
    \end{subfigure}
    \begin{subfigure}{0.\textwidth}
        \phantomsubcaption
        \label{fig:berry-phase}
    \end{subfigure}
    \begin{subfigure}{0.\textwidth}
        \phantomsubcaption
        \label{fig:adiabaticity}
    \end{subfigure}
\caption{Observation of a topological Berry phase via single-phonon interference. (a) The path on surface $\delta k_2$ enclosing the conical intersection as a function of $\{s_A,s_B\}$ and (b) not enclosing the conical intersection. The paths approximate contours of constant curvature (also shown in grey).  Surface $\delta k_2$ is mirror-symmetric about the plane indicated by the white dashed line. (c) Observation of $P_\downarrow$ as a function of evolution time after tuning the eigenmodes along an adiabatic, closed path in parameter space that encloses the conical intersection (gold disks) vs. a closed path that does not enclose the intersection (pink squares). The curves show a phase difference of $\Delta \phi = \pi \times(0.99 \pm0.01)$ as expected for the Berry phase acquired on the enclosing path. Solid lines are fits and error bars are $1\sigma$ of the mean. The eigenmodes were tuned with a transit duration of $T=$ 780 \textmu s. (d) The observed phase difference vs. the duration of traversing the enclosing and non-enclosing paths. The solid line is the phase difference predicted by numerical integration of the Schr{\"o}dinger equation of the 2D-array for the enclosing and non-enclosing paths in parameter space. Error bars are 68~\% confidence intervals obtained from 5000 bootstrapping trials.}
\end{figure*}
The precise controllability of our system enables us to prepare a superposition of states and perform interferometry between them to reveal the Berry phase with high contrast. When setting $s_A(0)=s_B(0)=1/(\alpha-1)$ and injecting a single phonon into ion C, this results in the state
\begin{equation}
\ket{\psi(0)} =\ket{001}_{ABC} =\frac{1}{\sqrt{2}} \left(\ket{010}_{123}+ \ket{001}_{123}\right),
\end{equation}
an equal amplitude superposition of eigenstates (single phonon Bell state) on the top two manifolds (see \nameref{sec:methods} for details). Then, $s_{A,B}(t)$ are changed along a path of approximately constant $\delta k_2$ ($\delta k_2(s_A(t),s_B(t))=\delta k_2(s_A(0),s_B(0))$ for all $t$), as shown in pink in Figs.~\ref{fig:berry-enclosed-path} and \ref{fig:berry-non-enclosed-path}. When the halfway point at  $s_A(T/2)=s_B(T/2)=\tfrac{\sqrt{2}-1-(\sqrt{2}-3)\alpha}{\alpha^2-1}$ is reached, the path shown in Fig. \ref{fig:berry-non-enclosed-path} turns around and doubles back on itself while the path shown in Fig.~\ref{fig:berry-enclosed-path} continues on along the mirror image of the first part ($s_{A,B}(t+T/2)=s_{B,A}(t),~0\leq t \leq T/2$) and encloses the conical intersection. Due to the symmetry of the system, all eigenvalue surfaces obey $\delta k_j(s_A,s_B)=\delta k_j(s_B,s_A)$, and therefore the enclosing and non-enclosing paths pick up equal dynamical phases. The component traveling on the largest eigenvalue surface never encloses a conical intersection and thus serves as a reference. After returning to the initial $s_A$ and $s_B$ values at $T$, the state at $T+t$ with $t>0$ has evolved to 
\begin{equation}
\ket{\psi(T+t)} = \frac{1}{\sqrt{2}}  \left(e^{i( \delta \omega_2 t+\phi_2+\phi_\mathrm{Berry})}\ket{010}_{123} + e^{i (\delta \omega_3 t+\phi_3)}\ket{001}_{123}\right).
\end{equation}
Here $\phi_{2}$ and $\phi_{3}$ are the dynamical phases picked up by the two eigenstates in the superposition on their trajectories (designed to be the same for the encircling and non-encircling paths) and $\phi_\mathrm{Berry}= \pi$ is the phase acquired by the $\ket{010}_{123}$ eigenstate from encircling the conical intersection adiabatically; for the path where the intersection was not encircled, $\phi_\mathrm{Berry}=0$.
When determining $P_{001}(T+t)$ as in the previously described experiments after traversing the paths in $T=$ 780 \textmu s, the Berry phase produces a measured phase shift between the interference fringes shown in Fig. \ref{fig:berry-phase} of $\Delta \phi =(0.99\pm 0.01)\pi$, determined by fitting each fringe to a single sinusoid with frequency, phase and amplitude as free parameters and extracting their phase difference from the fit parameters. We verify the independence of this phase from the shape of the path by implementing the same scheme for several other paths (see Supplementary Information, Figs.~\ref{fig:methods-berry-larger}--\ref{fig:methods-berry-loops}). When reducing the duration $T$ for traversing the pair of paths shown in Figs. \ref{fig:berry-enclosed-path} and \ref{fig:berry-non-enclosed-path} we observe that the relative phase is reduced from $\pi$ in the near-adiabatic case, $T \gtrsim $ 800 \textmu s to near zero for $T< 200$ \textmu s, as shown by the data points in Fig. \ref{fig:adiabaticity}; the solid green line indicates the phase difference predicted from numerically solving the Schr{\"o}dinger equation for the enclosing and non-enclosing paths. Non-adiabatic evolution also changes the contrast of the interference pattern (not shown) since transitions to the third eigenmode are no longer negligible. 
\section{Discussion and Outlook}\label{sec:discussion}
In summary, we precisely prepare and control the quantum state of single phonons in coupled harmonic oscillators realized by the Coulomb interaction of three ions in a 2D microtrap array with full control of individual sites. We insert a single phonon into the array, study its behavior for static and dynamical site tuning and find close agreement with theoretical models of the ideal coupled quantum system. We observe a topological Berry phase by single phonon interference and characterize the behavior of the system when the adiabaticity of the process is gradually reduced.\\
\\
The long coherence times and precise control motivate future work with more complex quantum states involving multiple phonons as well as the internal degrees of freedom of the ions, possibly in larger arrays. The desired symmetries in interactions can be directly established by the array geometry and broken at will by individual tuning of the local potential wells the ions reside in. The wells can be tuned dynamically on timescales slower or faster than those of the phonon dynamics and studied on all length scales, from a single site to changing or probing properties of the whole array. Larger finite size arrays may also give rise to topological edge states that can be studied with single site resolution or enable studies of complex coupled bosonic degrees of freedom \cite{porras_2004b}, including lattice-gauge models \cite{bermudez_synthetic_2011} and boson sampling \cite{aaronson2013}. By also incorporating the internal (spin) degrees of freedom of ions, spin-boson models with precisely designed 2D geometries and dynamically tunable sites can be studied. Effective spin-spin interactions \cite{porras_2004,schmied_quantum_2011} can be simulated by driving the array with laser or microwave fields and adiabatically eliminating the ion motion. For example, geometrically frustrated spin models on triangular or Kagome lattices or spin glasses may be realized in this manner. All these systems can be studied with dynamic individual site control and resolution that allows one to introduce disorder or induce phase transitions.
\section{Acknowledgments and Statement of Work}\label{sec:acknowledgements}

We thank D.T.C. Allcock for the passively stable rf source used in the apparatus, S. Geller for discussions about statistical analysis, Raymond A. Haltli for contributions to the trap fabrication and D. Palani and R. Shaniv for reviewing the manuscript. We acknowledge funding from the NIST Quantum Initiative. Authors affiliated with the University of Colorado Boulder were supported by the CU-NIST PREP Program, operated jointly by the University of Colorado Boulder and NIST.\\
\\
Sandia National Laboratories is a multimission laboratory managed and operated by National Technology \& Engineering Solutions of Sandia, LLC, a wholly owned subsidiary of Honeywell International Inc., for the U.S. Department of Energy’s National Nuclear Security Administration under contract DE-NA0003525. This paper describes objective technical results and analysis. Any subjective views or opinions that might be expressed in the paper do not necessarily represent the views of the U.S. Department of Energy or the United States Government.\\
\\
J.F.N. and D.L. performed the experiments and analyzed the data, with N.K.L. assisting. K.C.M. and J.K. performed preliminary experiments and testing of the ion trap. C.W.H., M.G.B., and E.J.H. led the fabrication of the ion trap. R.S., R.J., and D.L. designed the trap's electrode geometry. S.L.T. and D.H.S., with assistance from R.J., designed and built the cryogenic ion trap system. D.J.W., A.C.W. and D.L. advised the scientific direction of the project during its development and, with D.H.S., secured its funding. D.H.S. provided technical guidance for apparatus and experimental design. J.F.N. and D.L. wrote the manuscript with input from all authors. D.L. supervised the project and developed the theoretical models of the experiments.
\newpage
\section{Methods}\label{sec:methods}
\subsection{Hessian and eigenmodes}
With the origin of the coordinate system at the center of an equilateral triangle of trapping sites A, B, C with mutual distance $d_0$ (see Fig. \ref{fig:trap}), the site coordinates are 
\begin{equation}\label{Eq:EquPos}
{\mathbf r}_A = 
\frac{d_0}{2 \sqrt{3}} \left(
\begin{array}{r}
-\sqrt{3}\\
 1\\
 0   
\end{array}
\right), ~~
{\mathbf r}_B = 
\frac{d_0}{\sqrt{3}} \left(
\begin{array}{r}
 0\\
 -1\\
 0   
\end{array}
\right),~~
{\mathbf r}_C = 
\frac{d_0}{2 \sqrt{3}} \left(
\begin{array}{r}
  \sqrt{3}\\
1\\
 0   
\end{array}
\right).
\end{equation}
We call the principal axes of the  rf pseudopotential wells around these sites with the smallest curvature the radial directions. They are tilted by $\theta \approx 19^{\rm o}$ out of the plane of the triangle and oriented as the following unit vectors (shown as purple arrows in Fig. \ref{fig:trap}, with $\hat{\mathbf{e}}_{RA}$ for site A): 
\begin{equation}\label{Eq:NorModVec}
\hat{{\mathbf e}}_{RA} = 
\frac{1}{2} \left(
\begin{array}{r}
 -\sqrt{3} \cos \theta\\
 \cos \theta\\
2 \sin \theta   
\end{array}
\right), ~~
\hat{{\mathbf e}}_{RB} = 
 \left(
\begin{array}{r}
0\\
- \cos \theta\\
\sin \theta   
\end{array}
\right),~~
\hat{{\mathbf e}}_{RC} =
\frac{1}{2} \left(
\begin{array}{r}
 \sqrt{3} \cos \theta\\
 \cos \theta\\
 2 \sin \theta   
\end{array}
\right).
\end{equation}
We assume the potential curvature along these radial axes to be identical for each well, with value $k_0$. Three identical ions with charge $q$ and mass $m$, one held at each site with equilibrium positions as specified in Eq. (\ref{Eq:EquPos}), will interact by their mutual Coulomb repulsion. Small displacements $(\delta_A, \delta_B, \delta_C)$ from the equilibrium positions, along the principal axes Eq. (\ref{Eq:NorModVec}) and taken to be much smaller than the ion-ion-spacing, change the modulus of the distance between the ions as
\begin{eqnarray}\label{Eq:IonDis}
d_{AB}  &=& |\mathbf{r}_A+\delta_A \hat{{\mathbf e}}_{RA}-\mathbf{r}_B-\delta_B \hat{{\mathbf e}}_{RB}|\nonumber, \\
d_{BC}  &=& |\mathbf{r}_B+\delta_B \hat{{\mathbf e}}_{RB}-\mathbf{r}_C-\delta_C \hat{{\mathbf e}}_{RC}|\nonumber, \\
d_{AC}  &=& |\mathbf{r}_A+\delta_A \hat{{\mathbf e}}_{RA}-\mathbf{r}_C-\delta_C \hat{{\mathbf e}}_{RC}|.
\end{eqnarray}
Assuming no displacement of the ions from their equilibrium positions normal to the respective axes in Eq. (\ref{Eq:NorModVec}), the potential energy of the Coulomb-coupled system as a function of the small displacements is given by
\begin{equation}
V_0 =\frac{1}{2}k_0\left(\delta_A^2+\delta_B^2+\delta_C^2 \right)+\frac{q^2}{4 \pi \epsilon_0}\left(\frac{1}{d_{AB}}+\frac{1}{d_{BC}}+\frac{1}{d_{AC}}\right), 
\end{equation}
where $\epsilon_0$ is the vacuum permittivity.
The Hessian matrix $\mathbf{H}_0$ composed of the second derivatives of $V_0$ with respect to $(\delta_A, \delta_B , \delta_C)$ with elements $(\mathbf{H}_{0})_{ik} = \frac{\partial^2 V_0}{\partial \delta_i \partial \delta_k}$, taken at $\delta_A=\delta_B=\delta_C=0$ can be rewritten as 
\begin{equation}\label{Eq:HesMatSup}
 \mathbf{H}_0=
k_{\rm offs}
\left(
\begin{array}{ccc}
  1&0&0 \\
  0&1&0 \\
  0&0&1   
\end{array}
\right)
+
\Delta k
\left(
\begin{array}{ccc}
  1&1&1 \\
  1&1&1 \\
  1&1&1   
\end{array}
\right) \equiv  k_{\rm offs} \mathbf{I} + \Delta k \mathbf{C}, 
\end{equation}
with the offset curvature
\begin{equation}
k_{\rm offs} = k_0 - \frac{q^2}{4 \pi \epsilon_0} \frac{9-15 \cos(2 \theta)}{8 d_0^3}>0,
\end{equation}
that multiplies the identity matrix $\mathbf{I}$  and is much larger than the pre-factor of the coupling matrix $\mathbf{C}$ in our system.
\begin{equation}\label{Eq:Delk}
\Delta k =  \frac{q^2}{4 \pi \epsilon_0} \frac{11+3 \cos(2 \theta)}{8 d_0^3}>0.
\end{equation}
Any vector is an eigenvector of the identity matrix and the coupling matrix $\Delta k\, \mathbf{C}$ has three identical rows, which implies that it has a null-space of dimension two (two non-zero eigenvectors with eigenvalue zero) and a non-zero eigenvalue $3\,\Delta k$. A certain set of  normalized eigenvectors and eigenvalues is
\begin{eqnarray}
{\mathbf c}^{(3)}=\frac{1}{\sqrt{3}}
\left(
\begin{array}{c}
  1 \\
  1 \\
  1   
\end{array}
\right),&
{\mathbf c}^{(1)}=\frac{1}{\sqrt{2}}
\left(
\begin{array}{r}
  0 \\
  1 \\
  -1   
\end{array}
\right),&
{\mathbf c}^{(2)}=\frac{1}{\sqrt{6}}
\left(
\begin{array}{r}
  2 \\
  -1 \\
  -1   
\end{array}
\right),\\
\delta k_3 = 3\, \Delta k,&\delta k_1=0,&\delta k_2=0.
\end{eqnarray}
but any linear combination $a~{\mathbf c}^{(1)} + b~{\mathbf c}^{(2)}$ is also an eigenvector with eigenvalue zero. Another choice of normalized eigenvectors with more obvious three-fold symmetry and complex components is
\begin{eqnarray}
{\mathbf c}^{(3)}=\frac{1}{\sqrt{3}}
\left(
\begin{array}{c}
  1 \\
  1 \\
  1   
\end{array}
\right), &
{\mathbf c}^{(+)}=\frac{1}{\sqrt{3}}
\left(
\begin{array}{r}
  1 \\
e^{i 2\pi/3} \\
e^{i 4\pi/3}
\end{array}
\right),&
{\mathbf c}^{(-)}=\frac{1}{\sqrt{3}}
\left(
\begin{array}{r}
 1 \\
 e^{-i 2\pi/3} \\
e^{-i 4\pi/3}
\end{array}
\right).
\nonumber\\
\delta k_3 = 3\, \Delta k,&\delta k_+=0,&\delta k_-=0.
\end{eqnarray}
In this set, exchanging sites leaves ${\mathbf c}^{(3)}$ unchanged. Exchanging sites cyclically (which corresponds to $\pm 2\pi/3$ rotations around the $z$-axis) alters the remaining eigenvectors by a global phase only and all other permutations transform ${\mathbf c}^{(+)} \leftrightarrow {\mathbf c}^{(-)}$  up to a global phase.\\
\\
The system can be tuned by applying small potential changes (on the order of 100 $\mu$V) that we call ``shims'' to the electrodes closest to the three ion sites, two of which are shown in purple in Fig.~\ref{fig:trap}. We neglect the very small static electric field at the positions of the ions caused by the shims and only consider the change in the well curvatures. The three-fold symmetry dictates that a change of curvature in well A $\Delta k\, s_A$ due to a potential applied to the electrode closest to site A results in equal changes of well curvature $\alpha \Delta k\ s_A$ at the two other sites B and C. In our system $\alpha \approx -0.383$. Moreover, a change of $\{s_A, s_B, s_C\}$ on the electrodes under A, B and C can be rewritten as a global change of all three electrodes by $k_{\rm offs}' = k_{\rm offs}+\Delta k\, s_C$ and local changes of $\{s_A-s_C, s_B-s_C,0\}$. Besides the relatively small eigenvalue change from $k_{\rm offs} \rightarrow k'_{\rm offs}$ with $|\Delta k\, s_C|\ll k_{\rm offs}$, the eigenmodes are unaffected by the global change. This renders the shim-parameter space effectively two-dimensional. The space can be explored by just changing potentials under sites A and B which results in curvature changes of $\Delta k\,\mathbf{S}(s_A, s_B)$ with the unit-free shim matrix
\begin{equation}
\mathbf{S}(s_A, s_B)= 
\left(
\begin{array}{ccc}
  s_A+\alpha s_B&0&0 \\
  0&s_B+\alpha s_A&0 \\
  0&0&\alpha(s_A+s_B)   
\end{array}
\right)
\end{equation}
The curvature changes add to $\mathbf{H}_0$ for the total Hessian matrix,
\begin{equation}\label{Eq:HesDet}
\mathbf{H}(s_A, s_B)=\mathbf{H}_0+\Delta k\, \mathbf{S}(s_A, s_B) =  k_{\rm offs} \mathbf{I} + \Delta k\, \mathbf{C} + \Delta k\,\mathbf{S}(s_A, s_B),
\end{equation}
which is identical to Eq.~(\ref{Eq:HesMat}) in the main text. The eigenvectors of $\mathbf{H}(s_A, s_B)$ are equal to the eigenvectors of $\mathbf{C} + \mathbf{S}(s_A, s_B)$ and the eigenvalues of $\mathbf{H}(s_A, s_B)$ can be written as $k_j \equiv k_{\rm offs}+\delta k_j$, with $j=\{0,1,2\}$. The general eigensystem with non-zero shims $s_A, s_B$ can be expressed analytically, but  only reduces to simple expressions in a few special cases.\\
\\
Due to the threefold symmetry of the unperturbed system, the eigenvalues are identical under exchange of $s_A$ and $s_B$, $k_j(s_A,s_B)=k_j(s_B,s_A)$. Moreover, when $s_A \neq 0, s_B=0$, wells B and C have equal curvatures that are different from the curvature in well A. The normal mode vector $\mathbf{c}^{(1)}=\tfrac{1}{\sqrt{2}}(0,-1,1)^T$, has no participation from the ion in site A and remains a normal mode vector, irrespective of the value of $s_A$. The eigenvalue of this decoupled mode changes linearly as $k = k_{\rm offs}+\Delta k\, \alpha\, s_A$. In analogy for $s_A=0,s_B\neq 0$ there is always a mode with no participation in B. When $s_A = s_B \neq 0$, sites A and B have equal curvatures that are different from the curvature of site C, therefore one eigenvector has no participation in C. The eigenvalues $\delta k_j(s_A,s_B)$ are shown in Fig. \ref{fig:mode-surfaces} as a function of shim parameters $\{s_A,s_B\}$. The eigenvalues form three surfaces that obey the symmetry $\delta k_j(s_A,s_B)=\delta k_j(s_B,s_A)$ and $\delta k_1$ and $\delta k_2$ have a conical intersection at $\{s_A,s_B\} =\{0,0\}$ where two eigenvalues are degenerate and equal to zero. The $\delta k_j$ are ordered by their magnitude, $\delta k_1 \leq 0 \leq \delta k_2 < \delta k_3$ and shown in pink, blue, and green, respectively.
\subsection{Quantum states in the uncoupled basis and the eigenmode basis}
We can use basis states of the radial modes composed of products of number states of the three uncoupled harmonic oscillators situated at sites A, B, C with notation $\ket{klm}_{ABC}$, where the $k,l,m$ are equal to zero or positive integers.  This is a complete basis and an arbitrary state at time $t=0$ can be written as
\begin{equation}\label{Eq:ProSta}
    \ket{\psi(t=0)}_{ABC} = \sum_{klm} u_{klm} \ket{klm}_{ABC} = \sum_{klm} \frac{u_{klm}}{\sqrt{k!\,l!\, m!}}(\hat{a}^\dag_A)^k(\hat{a}^\dag_B)^l(\hat{a}^\dag_C)^m \ket{000}_{ABC}.
\end{equation}
In analogy, we denote the products of number states in the coupled radial mode eigenbasis as $\ket{klm}_{123}$. The ground state is identical in any basis, $\ket{000}_{ABC} = \ket{000}_{123}$. When $k_{\rm offs}\gg \Delta k$, the coupled radial eigenmode frequencies can be related to the curvatures as
\begin{eqnarray}\label{Eq:DelOme}
\omega_j = \sqrt{\frac{k_{\rm offs}+\delta k_j}{m}} &\approx& \omega_R+\delta \omega_j,\nonumber \\
\omega_{R}=\sqrt{\frac{k_{\rm offs}}{m}}, && \delta \omega_j = \frac{\delta k_j}{2 m\,\omega_R}.
\end{eqnarray}
The time evolution of the eigenmode creation operators $\hat{a}^\dag_j$ ($j\epsilon \{1,2,3\})$ is particularly simple,
\begin{equation}
\hat{a}^\dag_j(t) = \hat{a}^\dag_j(0)e^{i (\omega_R+\delta \omega_j)t}.
\end{equation}
The creation operators in the uncoupled basis can be rewritten in terms of the creation operators in the coupled eigenbasis by inserting the eigenmode coordinates of $\mathbf{H}$ in the notation $\mathbf{c}^{(k)}=(c^{(k)}_A,c^{(k)}_B,c^{(k)}_C)^T$
\begin{eqnarray}
\hat{a}^\dag_A(t) =e^{i \omega_R t}\left( c^{(1)}_A\hat{a}^\dag_1(0)e^{i \delta \omega_1 t}+c^{(2)}_A\hat{a}^\dag_2(0)e^{i \delta \omega_2 t}+c^{(3)}_A\hat{a}^\dag_3(0)e^{i \delta \omega_3 t}\right)\nonumber\\
\hat{a}^\dag_B(t) =e^{i \omega_R t}\left( c^{(1)}_B\hat{a}^\dag_1(0)e^{i \delta \omega_1 t}+c^{(2)}_B\hat{a}^\dag_2(0)e^{i \delta \omega_2 t}+c^{(3)}_B\hat{a}^\dag_3(0)e^{i \delta \omega_3 t}\right)\nonumber\\
\hat{a}^\dag_C(t) =e^{i \omega_R t}\left( c^{(1)}_C\hat{a}^\dag_1(0)e^{i \delta \omega_1 t}+c^{(2)}_C\hat{a}^\dag_2(0)e^{i \delta \omega_2 t}+c^{(3)}_C\hat{a}^\dag_3(0)e^{i \delta \omega_3 t}\right)
\end{eqnarray}
This implies that $\ket{\psi(t)}_{ABC}$ written in terms of the coupled eigenbasis creation operators will contain terms that are polynomial in the creation operators
\begin{equation}
[\hat{a}^\dag_A(t)]^k\ket{000}_{ABC} =e^{ik\omega_R t}\left[ c^{(1)}_A\hat{a}^\dag_1(0)e^{i \delta \omega_1 t}+c^{(2)}_A\hat{a}^\dag_2(0)e^{i \delta \omega_2 t}+c^{(3)}_A\hat{a}^\dag_3(0)e^{i \delta \omega_3 t}\right]^k \ket{000}_{123}.
\end{equation}
The sum of all these terms can lead to multi-path interference as a function of time, which can give rise to, for example, the complexity in boson sampling \cite{aaronson2013}. However, all terms in the polynomial will be proportional to $[\hat{a}^\dag_1(0)]^{k'}[\hat{a}^\dag_2(0)]^{l'}[\hat{a}^\dag_3(0)]^{m'}$ where the sum of exponents is preserved $k'+l'+m'=k$. This is also true for the other polynomials with $l$ and $m$ excitations in Eq.(\ref{Eq:ProSta}), meaning that the total number of excitations in an initial product of number states $\ket{klm}_{ABC}$ is conserved and equal to $k+l+m$ at all times.\\
\\
The time dependence simplifies for initial states with a single excitation, as the ones prepared in our experiments. In particular the state $\ket{\psi(0)} = \ket{001}_{ABC}$ evolves to  
\begin{equation}\label{Eq:TimSta}
    \ket{\psi(t)} = e^{i \omega_R t}\left( c^{(1)}_Ce^{i \delta \omega_1 t} \ket{100}_{123}+c^{(2)}_C e^{i \delta \omega_2 t}\ket{010}_{123}+c^{(3)}_C e^{i \delta \omega_3 t}\ket{001}_{123}\right),
\end{equation}
the expression stated as Eq.~(\ref{Eq:TimEvo}) in the main text. The probability $P_{001}$ of finding the phonon in site C at a later time $t>0$ consists of a constant and up to three oscillating terms at the differences of mode eigenfrequencies
\begin{eqnarray}\label{Eq:PhoInt}
P_{001}=\left|_{ABC}\langle 001 \ket{\Psi(t)}\right|^2& =& \left|c^{(1)}_C\right|^4 + \left |c^{(2)}_C\right|^4 + \left|c^{(3)}_C\right|^4\nonumber\\
&&+2 \left|c^{(1)}_C\right|^2\left|c^{(2)}_C\right|^2 \cos[(\delta \omega_2-\delta \omega_1)t]\nonumber\\
&&+ 2\left|c^{(1)}_C\right|^2\left|c^{(3)}_C\right|^2 \cos[(\delta \omega_3-\delta \omega_1)t]\nonumber\\
&&+2 \left|c^{(2)}_C\right|^2\left|c^{(3)}_C\right|^2 \cos[(\delta \omega_3-\delta \omega_2)t],
\end{eqnarray}
where certainty is reached whenever all three cosines are equal to one (for example when $t=0$), then  $P_{001}=\left( \left|c^{(1)}_C\right|^2 + \left|c^{(2)}_C\right|^2 + \left|c^{(3)}_C\right|^2\right)^2 =1$. If any of the third components $\{c^{(1)}_C,c^{(2)}_C,c^{(3)}_C\}$ of the three coupled basis eigenvectors is equal to zero, or if the frequencies of two eigenmodes are identical $\delta\omega_k=\delta\omega_l$ for $k\neq l$, only one time dependent term remains.
\subsection{Observing a Berry phase}
In the special case where $s_{A0} = s_{B0} = 1/(\alpha-1)$ the eigensystem becomes
\begin{eqnarray}\label{Eq:BerIni}
{\mathbf c}^{(1)}=\tfrac{1}{\sqrt{2}}
\left(
\begin{array}{r}
 -1 \\
  1 \\
  0  
\end{array}
\right),&
{\mathbf c}^{(2)}=\tfrac{1}{2}
\left(
\begin{array}{r}
  -1 \\
  -1 \\
  \sqrt{2} 
 \end{array}
\right),&
{\mathbf c}^{(3)}=\tfrac{1}{2}
\left(
\begin{array}{c}
  1 \\
  1 \\
  \sqrt{2}   
\end{array}
\right),\nonumber\\
\delta k_1=\Delta k \tfrac{\alpha+1}{\alpha-1},& \delta k_2= \Delta k(3-\sqrt{2}+\tfrac{2}{\alpha-1}),&\delta k_3=\Delta k(3+\sqrt{2}+\tfrac{2}{\alpha-1}). \nonumber\\
\end{eqnarray}
The first eigenmode does not participate in site C, while the other two eigenmodes participate equally. When injecting a phonon into site C the phonon state at $t=0$ is a single phonon entangled (Bell) state in the coupled eigenbasis
\begin{equation}
\ket{\psi(0)} = \ket{001}_{ABC}= \frac{1}{\sqrt{2}} \left(\ket{010}_{123}+ \ket{001}_{123}\right).
\end{equation}
The two independent parameters $\{s_A, s_B\}$ that change the external potential curvatures can now be tuned sufficiently slowly compared to $2 \pi/|\delta \omega_k -\delta \omega_l|$ for $\{k,l\}\, \epsilon \{ 1,2,3\}$,  $k\neq l$ such that the adiabatic theorem can be applied and the eigenstates of the system approximately remain eigenstates throughout the tuning of $\{s_A, s_B\}$. If a closed path in parameter space is traversed between $t=0$ and $t=T$, $\{s_A(T), s_B(T)\} =\{s_A(0), s_B(0)\}=\{s_{A0},s_{B0}\}$, the initial and final eigenstates only differ by a phase and the final state can be written as
\begin{equation}\label{Eq:BerSta}
\ket{\psi(T)}= \frac{1}{\sqrt{2}} \left(e^{i \phi_2} e^{i \phi_{\rm Berry}}\ket{010}_{123}+ e^{i \phi_3}\ket{001}_{123}\right),
\end{equation}
where
\begin{equation}
    \phi_{2,3}=\int_0^T dt' \delta \omega_{2,3}(s_A(t'),s_B(t'))
\end{equation}
is the dynamical phase picked up due to the instantaneous frequencies $\delta \omega_{2,3}(s_A(t'),s_B(t'))$ of the eigenmodes 2 and 3 along the path. The dynamical phases are time dependent generalizations of the dynamical phases $\phi_{2,3}=\delta \omega_{2,3} t$ that appear in Eq.~(\ref{Eq:TimSta}) for $s_A,s_B$ constant in time. In addition, each eigenstate can pick up a Berry phase $\phi_{\rm Berry}$ that does not depend on details of the path but rather on whether or not the path encloses the conical intersection at $\{s_A,s_B\}=\{0,0\}$~\cite{berry_phase_1984}. The surface $\delta \omega_3(s_A,s_B)$ has no conical intersection, so the component proportional to $\ket{001}_{123}$ will not pick up a Berry phase, irrespective of the path. For the component proportional to $\ket{010}_{123}$, if the intersection is enclosed adiabatically, $\phi_{\rm Berry} = \pi$,
otherwise $\phi_{\rm Berry} = 0$.\\
\\
To isolate the Berry phase, it is advantageous to pick pairs of paths that accumulate the same dynamical phases while one path encloses the conical intersection and the other does not. Due to the symmetry of the eigenvalue surfaces, there is a second point $ s_{A1}=s_{B1}=\tfrac{\sqrt{2}-1-(\sqrt{2}-3)\alpha}{\alpha^2-1} $ with equal curvature as the initial point, $\delta k_2(s_{A0},s_{B0})=\delta k_2(s_{A1},s_{B1})$ and $\delta k_3(s_{A0},s_{B0})=\delta k_3(s_{A1},s_{B1})$, and these points can be connected by paths of constant curvatures $\delta k_2(s_A,s_B)=\delta k_2(s_{A0},s_{B0})$ and $\delta k_3(s_A,s_B)=\delta k_3(s_{A0},s_{B0})$, respectively, as shown in Figs.~\ref{fig:berry-enclosed-path} and \ref{fig:berry-non-enclosed-path}. A path that starts at $s_{A0}=s_{B0}$, turns around at $s_{A1}=s_{B1}$, doubles back onto itself and returns to $s_{A0}=s_{B0}$ will not enclose the conical intersection (Fig. \ref{fig:berry-non-enclosed-path}). Since the eigenvalues are identical under exchange of the arguments, $\delta \kappa_j(s_A,s_B)=\delta \kappa_j(s_B,s_A)$, this path will pick up equal dynamical phases as the path that continues clockwise around the intersection shown in Fig. \ref{fig:berry-enclosed-path} and returns to the starting point. Depending on which path is chosen, a Berry phase of zero or $\pi$ is picked up and the difference will manifest itself as a phase shift of $\pi$ in the probability $P_{001}(T+t)$ ($t>0$) in Eq.~(\ref{Eq:PhoInt}) to find the phonon at site C for different delays after either one of the paths has been traversed.\\
\\
We explore a number of different pairs of paths with equal dynamical phases, with larger or smaller areas, with more changes of directions and even connecting a number of smaller loops. In all cases, we get a relative phase close to $\pi$ between the interference traces recorded after traversing the respective paths (Figs. \ref{fig:methods-berry-smaller}--\ref{fig:methods-berry-loops}). If the paths are traversed so rapidly that the adiabatic approximation breaks down, the acquired topological phase is gradually reduced from $\pi$ to zero (Fig. \ref{fig:adiabaticity}). The latter limit can be understood by considering an extremely fast cyclic change of $\{s_A,s_B\}$. In this case, the system dynamics are too slow to react substantially to the changes and it approximately remains in its initial state, irrespective of the chosen path.
\subsection{Fitting the data in Fig.~\ref{fig:spectrum}}
A set of 12 data points spanning approximately $\delta f= \pm 2$ kHz around each of the observed resonance peaks in Fig.~\ref{fig:spectrum} for fixed $\delta V_A$ and varying excitation frequency $f_R+\delta f$ was fitted to a Gaussian that yielded the center excitation frequency difference $\delta f_0$ and the width of the Gaussian fit. Resonances close to the crossing ($\delta V_A =\{-0.5, 0, 0.5\}$ mV) where two of the peaks are unresolved in the data were removed, because their convolution produces a single peak with a position that does not easily yield the frequencies of the two underlying resonances. The remaining peak centers approximately form three lines along the light-colored regions in Fig.~\ref{fig:spectrum}. The model for the eigenmode frequencies is derived from $\mathbf{H}(s_A, s_B)$ by substituting 
$$
k_{\rm offs} \rightarrow f_R =\tfrac{1}{2 \pi}\sqrt{k_{\rm offs}/m},~ \Delta k \rightarrow \Delta f = \tfrac{1}{2 \pi}\Delta k/(2 m \omega_R),~ s_A \rightarrow c\, \delta V_A,
$$
where $c$ is a constant that has units 1/V and describes the change in the uncoupled site frequency in units of $\Delta f$ when the shim potential is changed. The matrix
\begin{equation}
 \mathbf{F}=
f_R
\left(
\begin{array}{ccc}
  1&0&0 \\
  0&1&0 \\
  0&0&1   
\end{array}
\right)
+
\Delta f
\left(
\begin{array}{ccc}
  1+c\, \delta V_A&1&1 \\
  1&1+\alpha\,c\,\delta V_A&1 \\
  1&1&1+\alpha\,c\,\delta V_A   
\end{array}
\right),
\end{equation}
derived in this way has the same eigenvectors as the Hessian $\mathbf{H}$ and three eigenvalues
\begin{eqnarray}
    f_1 &=& f_R+ \Delta f\, \alpha\, c\, \delta V_A \\
    f_2 &=&f_R+ \tfrac{1}{2} \Delta f \left( 3  +c (1+\alpha) \delta V_A-\sqrt{9+2 c(\alpha-1) \delta V_A +c^2(\alpha-1)^2 \delta V_A}\right)\nonumber\\
    f_3 &=&f_R+ \tfrac{1}{2} \Delta f \left( 3  +c (1+\alpha) \delta V_A+\sqrt{9+2 c(\alpha-1) \delta V_A +c^2(\alpha-1)^2 \delta V_A}\right). \nonumber
\end{eqnarray}
Note that the frequency eigenvalues are not expressed and ordered in analogy to the curvature eigenvalue surfaces in the main text. The chosen form shows that $f_1$ can be directly fit to the set of experimental resonances that vary approximately linear in $\delta V_A$ to yield fitted values for $f_R = (3876.60 \pm 0.03)$ kHz (intercept) and $\Delta f\,\alpha\, c = (1.52 \pm 0.01)$ kHz/mV (slope). The remaining two resonance frequencies for equal $\delta V_A$ can be added to each other and the already determined quantities $2 f_R$ and $\Delta f\, \alpha\, c\, \delta V_A$ can be subtracted  
$$
f_2+f_3-\left(2 f_R+\Delta f\, \alpha\, c\, \delta V_A\right) = 3 \Delta f + \Delta f\, c\, \delta V_A, 
$$
which yields $\Delta f =(3.299 \pm 0.02)$ kHz (intercept) and $\Delta f\, c = (-3.97\pm 0.02)$ kHz/mV (slope). The isolated parameters can now be determined: $c= \Delta f\, c/\Delta f =(-1.202 \pm 0.009)$ 1/mV and $\alpha = \Delta f\, \alpha\, c/(\Delta f c) =(-0.383 \pm 0.003)$, where the uncertainties are derived by propagating the uncertainties of the fit parameters through the expressions stated above. By symmetry, $c$ and $\alpha$ also govern the change of frequencies when applying $\delta V_B \neq 0$; $s_{B}=c\, \delta V_{B}$. The frequency eigenvalues as a function of $\delta V_A$ as determined by the fit are plotted as solid lines in Fig.~\ref{fig:spectrum} and Fig.~\ref{fig:exchange-f} and are in excellent agreement with the data. The fitted function is plotted without any free parameters in Fig.~\ref{fig:exchange-f}. The same fit parameters are also used in the theoretical expressions of the bright population in Fig.s \ref{fig:exchange-b}--\ref{fig:exchange-e}. Additional fitted phonon exchange results not shown in \ref{fig:exchange-b}--\ref{fig:exchange-e} are displayed in Fig. \ref{fig:methods-all-three-ion-exchanges}.
The fitted values are also converted to curvatures according to Eq.(\ref{Eq:DelOme}) and used in Eq.(\ref{Eq:HesMat}) to plot the curvature surfaces in Fig. \ref{fig:mode-surfaces}, the intended paths for the Berry phase experiments in Fig.s \ref{fig:berry-enclosed-path}, \ref{fig:berry-non-enclosed-path} and the additional paths shown in the supplemental information.
\clearpage
\section{Supplemental Figures}\label{sec:methods-figs}
%
\renewcommand\thefigure{S\arabic{figure}}
\setcounter{figure}{0}
\begin{figure*}[htb!]
\centering
    \includegraphics[width=0.85\textwidth]{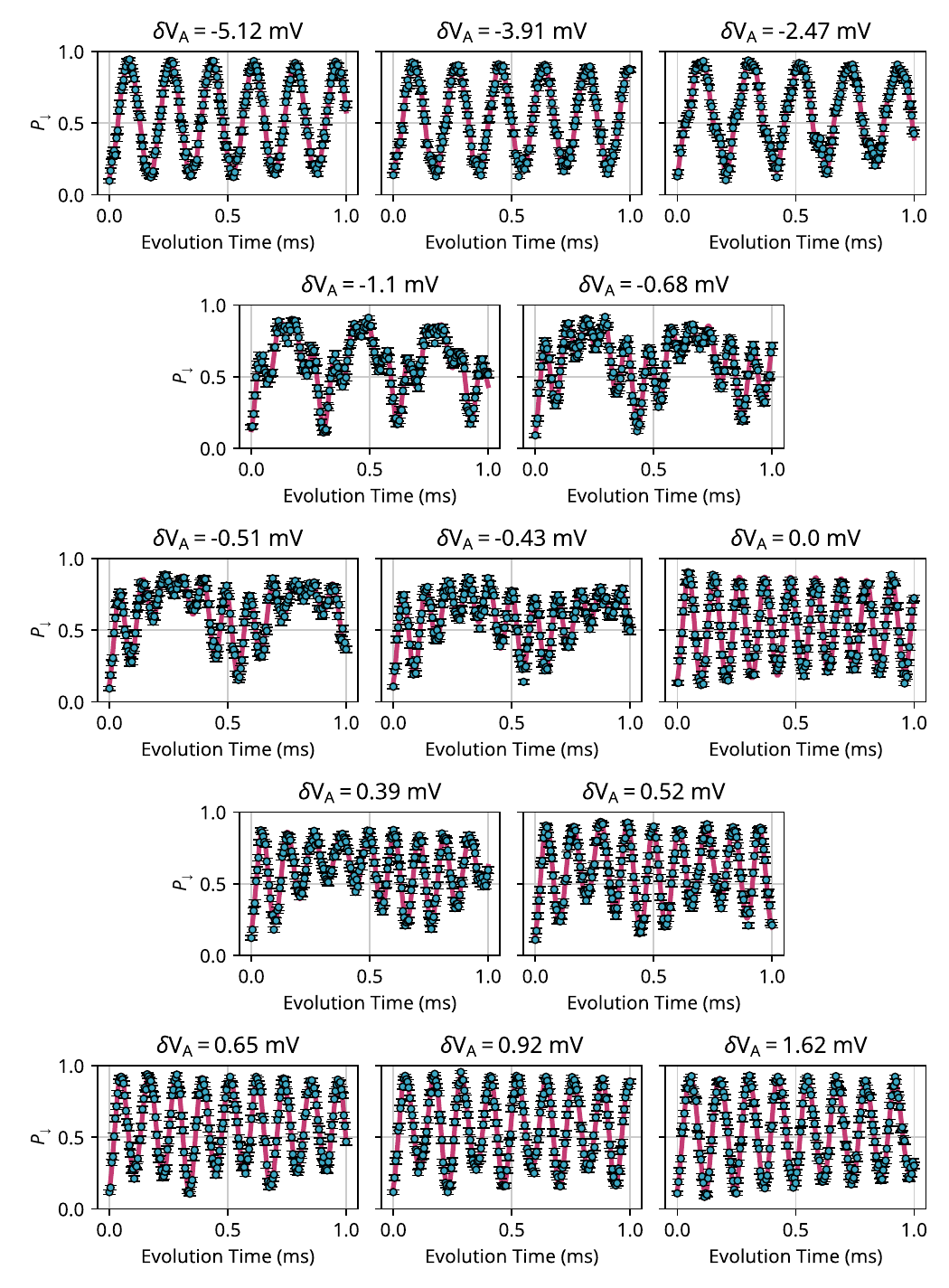}
    \caption{Single phonon interferences not shown in Figs. \ref{fig:exchange-b}-\ref{fig:exchange-e} for the remaining frequency data points as a function of $\delta \text{V}_\text{A}$ shown in Fig. \ref{fig:exchange-f}. As the voltage is changed slightly on the electrode under site A, the observed interference pattern changes. By fitting these data to a sum of sinusoids with different amplitudes (pink lines), the motional frequency differences of the system can be extracted and are shown in Fig. \ref{fig:exchange-f}. Error bars represent $1\sigma$ of $P_{\downarrow}$.}
    \label{fig:methods-all-three-ion-exchanges}    
\end{figure*}
\begin{figure*}[ht!]
\centering
    \begin{subfigure}{\textwidth}
        \centering
        \includegraphics[width=\textwidth]{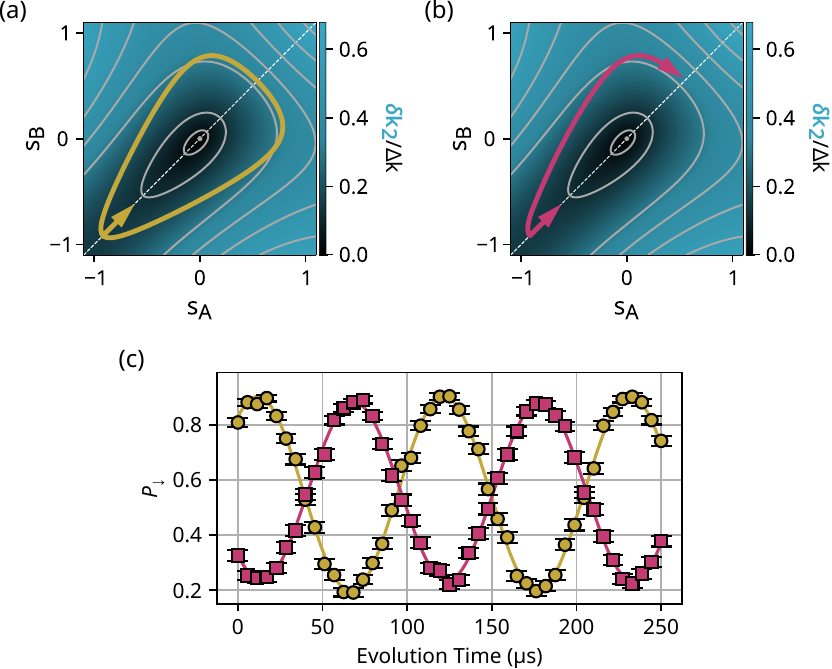}
        \phantomsubcaption
        \label{fig:methods-berry-larger-enclosed}
    \end{subfigure}
    \begin{subfigure}{0.\textwidth}
        \phantomsubcaption
        \label{fig:methods-berry-larger-non-enclosed}
    \end{subfigure}
    \begin{subfigure}{0.\textwidth}
        \phantomsubcaption
        \label{fig:methods-berry-larger-data}
    \end{subfigure}
\caption{Observation of a topological Berry phase via single-phonon interference. (a) The path on surface $\delta k_2$ as a function of $\{s_A,s_B\}$ enclosing the conical intersection but with a larger enclosed area than that of Fig. \ref{fig:berry-enclosed-path} and no longer traversing points of equal curvature. (b) The corresponding path not enclosing the conical intersection. Contours of constant curvature are shown in gray.  Surface $\delta k_2$ is mirror-symmetric about the plane indicated by the white dashed line. The eigenmodes were tuned with a transit duration of $T=$ 1.8 ms. (c) Observation of $P_\downarrow$ as a function of evolution time after tuning the eigenmodes along the closed path described above that encloses the conical intersection (gold disks) vs. the closed path that does not enclose the intersection (pink squares). The phase difference is $\Delta \phi = \pi \times(0.96 \pm0.02)$. Solid lines are fits and error bars are $1\sigma$ of the mean.}
\label{fig:methods-berry-larger}
\end{figure*}
\begin{figure*}[ht!]
\centering
    \begin{subfigure}{\textwidth}
        \centering
        \includegraphics[width=\textwidth]{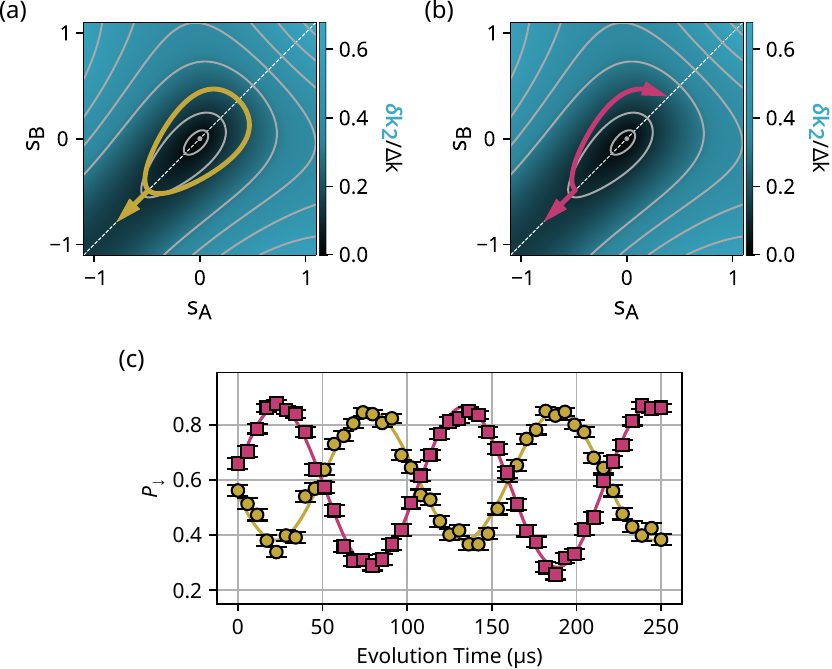}
        \phantomsubcaption
        \label{fig:methods-berry-smaller-enclosed}
    \end{subfigure}
    \begin{subfigure}{0.\textwidth}
        \phantomsubcaption
        \label{fig:methods-berry-smaller-non-enclosed}
    \end{subfigure}
    \begin{subfigure}{0.\textwidth}
        \phantomsubcaption
        \label{fig:methods-berry-smaller-data}
    \end{subfigure}
\caption{Observation of a topological Berry phase via single-phonon interference. (a) The path on surface $\delta k_2$ as a function of $\{s_A,s_B\}$ enclosing the conical intersection but with a smaller enclosed area than that of Fig. \ref{fig:berry-enclosed-path} and no longer traversing points of equal curvature. (b) The analogous path not enclosing the conical intersection. Contours of constant curvature are shown in gray.  Surface $\delta k_2$ is mirror-symmetric about the plane indicated by the white dashed line. The eigenmodes were tuned with a transit duration of $T=$ 1.8 ms. (c) Observation of $P_\downarrow$ as a function of evolution time after tuning the eigenmodes along the closed path described above that encloses the conical intersection (gold disks) vs. the closed path that does not enclose the intersection (pink squares). The phase difference is $\Delta \phi = \pi \times(0.97 \pm0.03)$. Solid lines are fits and error bars are $1\sigma$ of the mean. Unequal contrast is likely the result of insufficient adiabaticity of the paths.}
\label{fig:methods-berry-smaller}
\end{figure*}
\begin{figure*}[ht!]
\centering
    \begin{subfigure}{\textwidth}
        \centering
        \includegraphics[width=\textwidth]{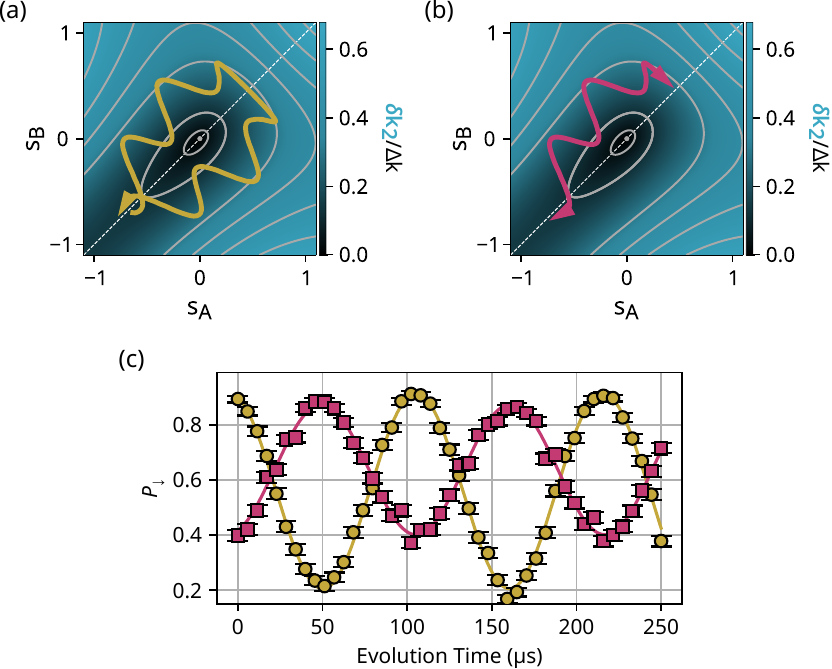}
        \phantomsubcaption
        \label{fig:methods-berry-wave-enclosed}
    \end{subfigure}
    \begin{subfigure}{0.\textwidth}
        \phantomsubcaption
        \label{fig:methods-berry-wave-non-enclosed}
    \end{subfigure}
    \begin{subfigure}{0.\textwidth}
        \phantomsubcaption
        \label{fig:methods-berry-wave-data}
    \end{subfigure}
\caption{Observation of a topological Berry phase via single-phonon interference. (a) The path on surface $\delta k_2$ as a function of $\{s_A,s_B\}$ enclosing the conical intersection oscillating about a path of near-equal curvature and (b) not enclosing the conical intersection. Contours of constant curvature are shown in gray.  Surface $\delta k_2$ is mirror-symmetric about the plane indicated by the white dashed line. The eigenmodes were tuned with a transit duration of 1.8 ms. (c) Observation of $P_\downarrow$ as a function of evolution time after tuning the eigenmodes along the paths described above. The path that encloses the conical intersection (gold disks) shows a phase difference of $\Delta \phi = \pi \times(1.02 \pm0.02)$ relative to the path that does not enclose the intersection (pink squares). Solid lines are fits and error bars are $1\sigma$ of the mean. Unequal contrast is likely the result of insufficient adiabaticity of the paths.}
\end{figure*}
\begin{figure*}[ht!]
\centering
    \begin{subfigure}{\textwidth}
        \centering
        \includegraphics[width=\textwidth]{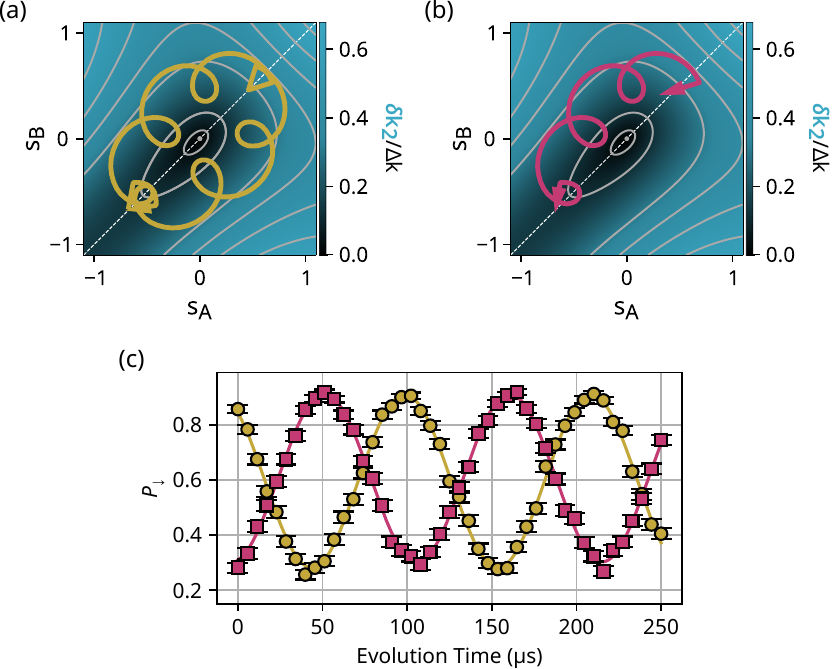}
        \phantomsubcaption
        \label{fig:methods-berry-loops-enclosed}
    \end{subfigure}
    \begin{subfigure}{0.\textwidth}
        \phantomsubcaption
        \label{fig:methods-berry-loops-non-enclosed}
    \end{subfigure}
    \begin{subfigure}{0.\textwidth}
        \phantomsubcaption
        \label{fig:methods-berry-loops-data}
    \end{subfigure}
\caption{Observation of a topological Berry phase via single-phonon interference. (a) The path on surface $\delta k_2$ as a function of $\{s_A,s_B\}$ enclosing the conical intersection with several additional windings.    (b) Path with equal dynamical phase but not enclosing the conical intersection. Contours of constant curvature are shown in gray.  Surface $\delta k_2$ is mirror-symmetric about the plane indicated by the white dashed line. The eigenmodes were tuned with a transit duration of 1.8 ms. (c) Observation of $P_\downarrow$ as a function of evolution time after tuning the eigenmodes along a closed path in parameter space shown in (a) that encloses the conical intersection (gold disks) vs. a closed path that does not enclose the intersection (shown in (b) pink squares). The observed phase difference is $\Delta \phi = \pi \times(0.86 \pm0.02)$. Solid lines are fits and error bars are $1\sigma$ of the mean. Deviation from the adiabatic value of $\Delta \phi = \pi$ are likely the result of too rapid changes of $\{s_A,s_B\}$ along the paths.}
\label{fig:methods-berry-loops}
\end{figure*}
\clearpage
\bibliography{references}
\end{document}